\documentclass[acus]{JAC2001}
\usepackage{graphicx}
\newcommand{\ud}{\mathrm{d}}
\begin{document}

\onecolumn 
\thispagestyle{empty}

\begin{center}
\begin{tabular}{p{130mm}}

\begin{center}
{\bf\Large 
NONLINEAR LOCALIZED COHERENT } \\
\vspace{5mm}

{\bf\Large SPECTRUM OF BEAM-BEAM INTERACTIONS}

\vspace{1cm}

{\bf\Large Antonina N. Fedorova, Michael G. Zeitlin}\\

\vspace{1cm}

{\bf\large\it
IPME RAS, St.~Petersburg, 
V.O. Bolshoj pr., 61, 199178, Russia}\\
{\bf\large\it e-mail: zeitlin@math.ipme.ru}\\
{\bf\large\it http://www.ipme.ru/zeitlin.html}\\
{\bf\large\it http://www.ipme.nw.ru/zeitlin.html}
\end{center}

\vspace{1cm}

\abstract{ 
We consider modeling for strong-strong beam-beam interactions
beyond preceding linearized/perturbative methods 
such as soft gaussian approximation
or FMM (HFMM) etc.
In our approach discrete coherent modes, discovered before, and
possible incoherent oscillations appear as
a result of  multiresolution/multiscale fast convergent decomposition 
in the bases of
high-localized exact nonlinear modes represented by wavelets or 
wavelet packets functions.
The constructed solutions represent the full multiscale spectrum
in all internal hidden scales from slow to  fast oscillating
eigenmodes. Underlying variational method provides algebraical control
of the spectrum.
}

\vspace{60mm}

\begin{center}
{\large Presented at the Eighth European Particle Accelerator Conference} \\
{\large EPAC'02} \\
{\large Paris, France,  June 3-7, 2002}
\end{center}
\end{tabular}
\end{center}
\newpage

\title{NONLINEAR LOCALIZED COHERENT SPECTRUM OF BEAM-BEAM INTERACTIONS}
\author{Antonina N. Fedorova, Michael G. Zeitlin \\
IPME, RAS, St.~Petersburg, 
V.O. Bolshoj pr., 61, 199178, Russia
\thanks{e-mail: zeitlin@math.ipme.ru}\thanks{http://www.ipme.ru/zeitlin.html;
http://www.ipme.nw.ru/zeitlin.html}}

\maketitle

\begin{abstract}

We consider modeling for strong-strong beam-beam interactions
beyond preceding linearized/perturbative methods 
such as soft gaussian approximation
or FMM (HFMM) etc.
In our approach discrete coherent modes, discovered before, and
possible incoherent oscillations appear as
a result of  multiresolution/multiscale fast convergent decomposition 
in the bases of
high-localized exact nonlinear modes represented by wavelets or 
wavelet packets functions.
The constructed solutions represent the full multiscale spectrum
in all internal hidden scales from slow to  fast oscillating
eigenmodes. Underlying variational method provides algebraical control
of the spectrum.

\end{abstract}

\section{INTRODUCTION}

We consider the first steps of analysis
of beam-beam interactions in some collective model approach.
It is well-known that neither direct PIC modeling
nor soft-gaussian approximation provide 
reasonable resolution of computing time/noise problems
and understanding of underlying complex nonlinear dynamics [1], [2].
Recent analysis, based as on numerical simulation as on 
modeling, demonstrates 
that presence of coherent modes inside the spectrum leads to 
oscillations and growth of beam transverse size and deformations of 
beam shape. This leads to strong limitations for operation of LHC.
Additional problems appear as a result of continuum spectrum of
incoherent oscillations in each beam.
The strong-strong collisions of two beams also lead to
variation of transverse size. According to [2]  it is reasonable to find 
nonperturbative solutions at least in the important particular cases. 
Our approach based on wavelet analysis technique is in some sense 
the direct generalization of Fast Multipole Method (FMM) and related approaches (HFMM).
After set-up based on Vlasov-like model (according [2]) in part 2, 
we consider 
variational-wavelet approach [3]-[17] in framework of powerful technique based 
on Fast Wavelet Transform (FWT) operator representations [18] in section 3.
As a result we have  multiresolution/multiscale fast convergent decomposition 
in the bases of
high-localized exact nonlinear eigenmodes represented by wavelets or 
wavelet packets functions.
The constructed solutions represent the full multiscale spectrum
in all internal hidden scales from slow to  fast oscillating
eigenmodes. Underlying variational method provides algebraical control
of the spectrum.

\section{VLASOV MODEL FOR BEAM-BEAM INTERACTIONS}

Vlasov-like equations describing evolution of the phase space distributions
$\psi^j=\psi^j(x,p_x,\theta)$ ($j=1,2$) for each beam are [2]: 
\begin{eqnarray}
&&\frac{\partial\psi^j}{\partial\theta}=-q_xp_x\frac{\partial\psi^j}{\partial x}+\\
&&\Big(q_xx+\delta_p(\theta)4\pi\xi_xp.v.
\int^\infty_{-\infty}\frac{\rho^*(x',\theta)}{x-x'}\ud x'\Big)
\frac{\partial\psi^j}{\partial p_x}\nonumber
\end{eqnarray}
where
\begin{equation}
\rho^*(x,\theta)=\int^{\infty}_{-\infty}\psi^*(x,p_x,\theta)\ud p_x
\end{equation}
and $\psi^*$ is the density of the opposite beam, $q_x$ is unperturbed fractional tune, 
$\xi_x$ is horizontal beam-beam parameter,
$N$ is a number of particles,
$x$, $p_x$ are normalized variables.
This model describes 
horizontal oscillations of flat beams with one bunch per beam, 
one interaction point, equal energy,
population and optics for both beams.

\section{FWT BASED VARIATIONAL APPROACH}

One of the key points of wavelet approach demonstrates that for a large class of
operators wavelets are good approximation for true eigenvectors and the corresponding 
matrices are almost diagonal. FWT [18] gives  the maximum sparse form of operators
under consideration (1).
It is true also in case of our Vlasov-like system of equations (1). 
We have both differential 
and integral operators inside.
So, let us denote our (integral/differential) operator from equations (1)  
as  $T$ ($L^2(R^n)\rightarrow L^2(R^n)$) and its kernel as $K$.
We have the following representation:
\begin{equation}
<Tf,g>=\int\int K(x,y)f(y)g(x)\ud x\ud y
\end{equation}
In case when $f$ and $g$ are wavelets $\varphi_{j,k}=2^{j/2}\varphi(2^jx-k)$
(3) provides the standard representation for operator $T$.
Let us consider multiresolution representation
$
\dots\subset V_2\subset V_1\subset V_0\subset V_{-1}
\subset V_{-2}\dots
$. 
The basis in each $V_j$ is 
$\varphi_{j,k}(x)$,
where indices $\ k, j$ represent translations and scaling 
respectively. 
Let $P_j: L^2(R^n)\rightarrow V_j$ $(j\in Z)$ be projection
operators on the subspace $V_j$ corresponding to level $j$ of resolution:
$
(P_jf)(x)=\sum_k<f,\varphi_{j,k}>\varphi_{j,k}(x).
$ 
Let
$Q_j=P_{j-1}-P_j$ be the projection operator on the subspace $W_j$ ($V_{j-1}=V_j\oplus W_j$) then
we have the following "microscopic or telescopic"
representation of operator T which takes into account contributions from
each level of resolution from different scales starting with the
coarsest and ending to the finest scales [18]:
\begin{equation}
T=\sum_{j\in Z}(Q_jTQ_j+Q_jTP_j+P_jTQ_j).
\end{equation}
We remember that this is a result of presence of affine group inside this
construction.
The non-standard form of operator representation [18] is a representation of
operator T as  a chain of triples
$T=\{A_j,B_j,\Gamma_j\}_{j\in Z}$, acting on the subspaces $V_j$ and
$W_j$:
$
 A_j: W_j\rightarrow W_j, B_j: V_j\rightarrow W_j,
\Gamma_j: W_j\rightarrow V_j,
$
where operators $\{A_j,B_j,\Gamma_j\}_{j\in Z}$ are defined
as
$A_j=Q_jTQ_j, \quad B_j=Q_jTP_j, \quad\Gamma_j=P_jTQ_j.$
The operator $T$ admits a recursive definition via
\begin{eqnarray}
T_j=
\left(\begin{array}{cc}
A_{j+1} & B_{j+1}\\
\Gamma_{j+1} & T_{j+1}
\end{array}\right),
\end{eqnarray}
where $T_j=P_jTP_j$ and $T_j$ acts on $ V_j: V_j\rightarrow V_j$.
It should be noted that operator $A_j$ describes interaction on the
scale $j$ independently from other scales, operators $B_j,\Gamma_j$
describe interaction between the scale j and all coarser scales,
the operator $T_j$ is an "averaged" version of $T_{j-1}$.
We may compute such non-standard representations
for different operators (including Calderon-Zygmund or pseudodifferential). 
As in case of differential operator 
as in other cases
we need only to solve the system of linear algebraical
equations. 
The action of integral operator in equations (1) we may
consider as a Hilbert transform
\begin{equation}
(H\rho^*)(x)=\frac{1}{\pi} p.v.\int^\infty_{-\infty}
\frac{\rho^*(x',\theta)}{x'-x}\ud x'
\end{equation}
The representation of $H$ on $V_0$ is defined by the coefficients 
\begin{equation}
r_\ell=\int\varphi(x-\ell)(H\varphi)(x)\ud x,\quad \ell\in Z.
\end{equation}
which according to $FWT$ technique define also all other coefficients of 
the nonstandard representation. So we have $H=\{A_j,B_j,\Gamma_j\}_{j\in Z}$
with the corresponding matrix elements $a_{i-\ell}, b_{i-\ell}, c_{i-\ell}$
which can be computed from coefficients $r_\ell$ only:
\begin{eqnarray}
a_i&=&\sum^{L-1}_{k,k'=0}g_kg_{k'}r_{2i+k-k'}\nonumber\\
b_i&=&\sum^{L-1}_{k,k'=0}g_kh_{k'}r_{2i+k-k'}\\
c_i&=&\sum^{L-1}_{k,k'=0}h_kg_{k'}r_{2i+k-k'}\nonumber
\end{eqnarray}
The coefficients $r_\ell$ (7) can be obtained from
\begin{equation}
r_\ell=r_{2\ell}+\sum^{L/2}_{k=1}d_{2k-1}(r_{2\ell-2k+1}+
r_{2\ell+2k-1})
\end{equation}
where $d_n$ are the so called autocorrelation coefficients of 
the corresponding quadratic mirror filter $\quad \{h_k\}^{L-1}_{k=0}$:
$\qquad d_n=2\sum^{L-1-n}_{i=0}h_ih_{i+n}$, $ n=1,\dots,L-1,$
$d_{2k}=0$, $k=1,\dots,L/2-1$, 
$g_k=(-1)^kh_{L-k-1}$, $k=0,\dots,L-1$,
which parametrizes the basic refinement equation
$
\varphi(x)=\sqrt{2}\sum^{L-1}_{k=0}h_k\varphi(2x-k).
$
This equation really generates all wavelet zoo.
It is useful to add to the system (9) the following asymptotic condition 
$r_\ell=-1/\pi\ell+O(\ell^{-2M})$, which simplifies the solution procedure.
Then finally we have the following action of operator $T_j$
on sufficiently smooth function $f$:
\begin{equation}
(T_j f)(x)=\sum_{k\in Z}\left(2^{-j}\sum_{\ell}r_\ell f_{j,k-\ell}\right)
\varphi_{j,k}(x),
\end{equation}
in the wavelet basis $\varphi_{j,k}(x)=2^{-j/2}\varphi(2^{-j}x-k)$ where
\begin{equation}
f_{j,k-1}=2^{-j/2}\int f(x)\varphi(2^{-j}x-k+\ell)\ud x
\end{equation}
are wavelet coefficients. So, we have simple linear para\-met\-rization of
matrix representation of our operator (6) in wavelet bases
and of the action of
this operator on arbitrary vector in proper functional space.
The similar approach can be applied to other operators in (1).
Then we may apply our variational approach from [3]-[17].
Let L be an ar\-bit\-rary (non) li\-ne\-ar (dif\-fe\-ren\-ti\-al\-/\-in\-teg\-ral) 
operator corresponds to the system (1) with matrix dimension $d$, 
which acts on some set of functions
$\quad\Psi\equiv\Psi(\theta,x,p_x)=$\\*
$\Big(\Psi^1(\theta,x,p_x),\dots,$ $\Psi^d(\theta,x,p_x)\Big)$, 
$\quad \theta,x,p_x\in\Omega\subset{\bf R}^3$,\\*
$
L\Psi\equiv L(Q,\theta,x,p_x)\Psi(\theta,x,p_x)=0,
$
where
$Q\equiv Q_{d_1,d_2,d_3}(\theta,x,p_x,\partial /\partial \theta,\partial /\partial x,\partial /\partial p_x
,\int \ud x\ud p_x)$.
Let us consider now the N mode approximation for solution as the following ansatz (in the same way
we may consider different ansatzes) [17]:
\begin{equation}
\Psi^N(\theta,x,p_x)=\sum^N_{r,s,k=1}a_{rsk}A_r\otimes B_s\otimes C_k(\theta,x,p_x)
\end{equation}
We shall determine the coefficients of expansion from the following conditions
(different related variational approaches are considered in [3]-[16]):
$$
\ell^N_{k\ell m}\equiv\int(L\Psi^N)A_k(\theta)B_\ell(x)C_m(p_x)\ud \theta\ud x\ud p_x=0
$$
So, we have exactly $dN^3$ algebraical equations for  $dN^3$ unknowns $a_{rsk}$.
The solution is parametrized by solutions of two set of reduced algebraical
problems, one is linear or nonlinear
(depends on the structure of operator L) and the rest are some linear
problems related to computation of coefficients of algebraic equations.
These coefficients can be found  by some wavelet methods
by using
compactly supported wavelet basis functions for expansions (12).
We may consider also different types of wavelets including general wavelet packets.
The constructed solution has the following mul\-ti\-sca\-le\-/\-mul\-ti\-re\-so\-lu\-ti\-on 
decomposition via 
nonlinear high\--\-lo\-ca\-li\-zed eigenmodes 
{\setlength\arraycolsep{0pt}
\begin{eqnarray}\label{eq:z}
&&\psi(\theta,x,p_x)=\sum_{(i,j,k)\in Z^3}a_{ijk}A^i(\theta)B^j(x)C^k(p_x),\\
&&A^i(\theta)=A_N^{i,slow}(\theta)+\sum_{r\geq N}A^i_r(\omega_r\theta),\ \omega_r\sim 2^r \nonumber\\
&&B^j(x)=B_M^{j,slow}(x)+\sum_{l\geq M}B^j_l(k^1_lx),\ k^1_l\sim 2^l \nonumber\\
&&C^s(p_x)=C_L^{s,slow}(p_x)+\sum_{m\geq L}C^s_m(k^2_mp_x),\ k^2_m\sim 2^m \nonumber
\end{eqnarray}}
which corresponds to the full multiresolution expansion in all underlying time/space 
scales.
\begin{figure}[htb]
\centering
\includegraphics*[width=60mm]{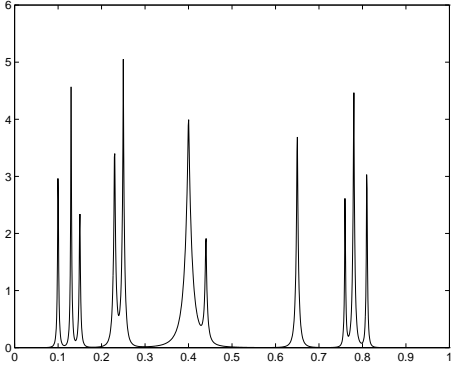}
\caption{Region of nonlinear resonances.}
\end{figure} 
\begin{figure}[htb]
\centering
\includegraphics*[width=60mm]{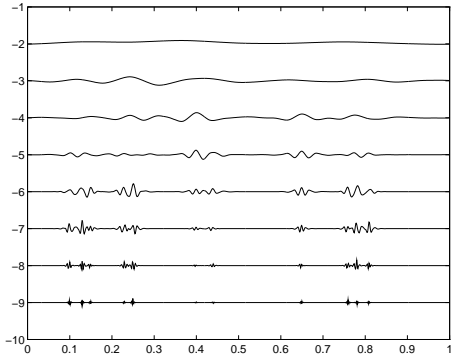}
\caption{Eigenmodes decomposition.}
\end{figure}  
Formula (\ref{eq:z}) gives us expansion into the slow part $f_{N,M,L}^{slow}$
and fast oscillating parts for arbitrary N, M, L.  So, we may move
from coarse scales of resolution to the 
finest one to obtain more detailed information about our dynamical process.
The first terms in the RHS of formulae (13) correspond on the global level
of function space decomposition to  resolution space and the second ones
to detail space.
The using of wavelet basis with high-localized properties provides fast convergence  
of constructed decomposition (13). 
In contrast with different approaches, formulae (13) does not use perturbation
technique or linearization procedures 
and represents dynamics via generalized nonlinear localized eigenmodes expansion.  
Numerical calculations are based on compactly supported
wavelets and related wavelet families.
Figures~1,2 demonstrate resonances region and corresponding 
nonlinear coherent eigenmodes decomposition
according to representation (13).

\end{document}